\documentclass[prd,aps,reprint,nobibnotes,twocolumn]{revtex4-1}

\usepackage{graphicx}
\usepackage{amsmath,amssymb}
\usepackage{braket}
\usepackage{commath}

\begin{document}

\title{Imaging with an ultra-thin reciprocal lens}

\author{Wenzhe Liu$^{1}$}
\email{e-mail: wliubh@connect.ust.hk (W. L.); jzi@fudan.edu.cn (J. Z.); phchan@ust.hk (C. T. C.)}
\author{Jingguang Chen$^{2}$}
\author{Tongyu Li$^{2}$}
\author{Zhe Zhang$^{2}$}
\author{Fang Guan$^{2}$}
\author{Lei Shi$^{2}$}
\author{Jian Zi$^{2}$}
\email{e-mail: wliubh@connect.ust.hk (W. L.); jzi@fudan.edu.cn (J. Z.); phchan@ust.hk (C. T. C.)}
\author{C. T. Chan$^{1}$}
\email{e-mail: wliubh@connect.ust.hk (W. L.); jzi@fudan.edu.cn (J. Z.); phchan@ust.hk (C. T. C.)}

\affiliation{$^{1}$ Department of Physics, The Hong Kong University of Science and Technology,  Hong Kong, China}
\affiliation{$^{2}$ State Key Laboratory of Surface Physics, Key Laboratory of Micro- and Nano-Photonic Structures (Ministry of Education), and Department of Physics, Fudan University, Shanghai 200433, China}

\begin{abstract}
  Imaging is of great importance in everyday life and various fields of science and technology. Conventional imaging is achieved by bending light rays originating from an object with a lens. Such ray bending requires space-variant structures, inevitably introducing a geometric center to the lens. To overcome the limitations arising from the conventional imaging mechanism, we consider imaging elements that employ a different mechanism, which we call reciprocal lenses. This type of imaging element relies on ray shifting, enabled by momentum-space-variant phase modulations in periodic structures. As such, it has the distinct advantage of not requiring alignment with a geometric center. Moreover, upright real images can be produced directly with a single reciprocal lens as the directions of rays are not changed. We realized an ultra-thin reciprocal lens based on a photonic crystal slab. We characterized the ray shifting behavior of the reciprocal lens and demonstrated imaging. Our work gives an alternative mechanism for imaging, and provides a new way to modulate electromagnetic waves.
\end{abstract}

\maketitle


Lenses are of fundamental importance in modern society. They can produce images and enable us to see objects from bacteria to distant stars \cite{jenkins1976fundamentals,born1999principles,hecht2017optics}. Animal species in every major phylum have evolved to make lenses from biological light-bending structures. Artificial lenses, invented about 3000 years ago, have played critical roles in almost every field of science and technology, from semiconductor device fabrication \cite{ito2000pushing,biswas2012advances} to observation of the cosmos \cite{hester1996hubble,nan2011five}. In the past decade, a new kind of lens, which modulates electromagnetic waves using many assorted subwavelength resonators has been realized \cite{khorasaninejad2016metalenses,wang2018broadband,li2018achromatic,schlickriede2020nonlinear,fan2022experimental,fu2022steerable,hua2022ultra,moon2022tutorial}. Called ``meta-lenses'', these lenses have several advantages over conventional lenses such as being far less bulky. These lenses still have some limitations. A lens has a geometric center, which means that it must be well aligned to produce an image of good quality. Moreover, a single lens will obey the Gaussian lens formula and can only produce inverted real images \cite{jenkins1976fundamentals,born1999principles,hecht2017optics}. Therefore, a system of lenses must be applied to obtain upright real images. In this work, we will introduce another class of lenses, ``reciprocal lenses''. Reciprocal lenses follow a distinct lens formula and have a different imaging mechanism. A single reciprocal lens has no geometric center, and can produce upright real images. We report our experimental realization of an ultra-thin reciprocal lens in the microwave range and verify the imaging mechanism.

\section{Revisiting conventional lenses}
\begin{figure*}[htpb]
\includegraphics[scale=0.9]{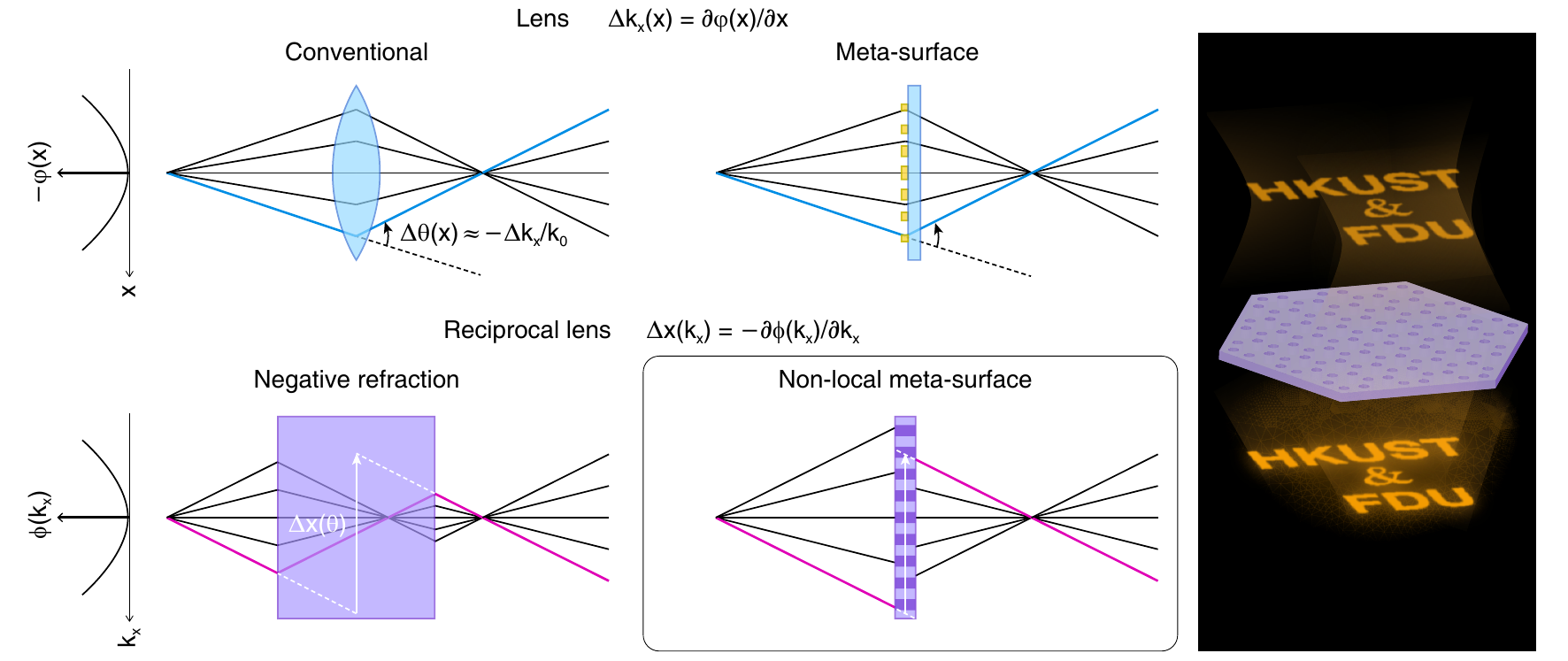}
\centering
\caption{Schematics of the proposed imaging mechanism compared to existing mechanisms. From the viewpoint of ray optics, imaging is realized by bending rays at different positions. Conventional lenses bend rays by refraction using a curved high refractive index medium. “Meta-lenses” bend rays by using meta-surfaces carrying sub-wavelength resonators and are thinner than bulky conventional lenses. All these lenses produce inverted real images. There is another class of lenses, which we call “reciprocal lens”. Instead of bending rays, they produce images by angle-dependent shifting of rays, as illustrated. Bulky reciprocal lenses can be realized in theory based on negative refraction. Here, we report our experimental realization of an ultra-thin reciprocal lens by implementing a photonic crystal slab as a non-local meta-surface. One particular ray for each mechanism has been highlighted to show the difference. As shown, the image produced by our reciprocal lens is upright but real.}
\label{fig:1}
\end{figure*}

We first revisit the concept and imaging mechanism of conventional lenses before we introduce the notion of ``reciprocal lenses''. Only the far-field propagation of waves will be considered, and we take the viewpoint of ray optics for ease of understanding. In a nutshell, an ideal lens focuses electromagnetic waves by bending rays differently at different positions in the lens plane, as illustrated in Fig. \ref{fig:1}. The lens is assumed to be cylindrically symmetric about the $z$ axis for simplicity and we reduce the system to a two dimensional problem in the $x$-$z$ plane for now. For each ray intersecting the lens plane at $x$, the angle of propagation would be changed by $\Delta \theta (x)$. In the paraxial limit, the angle change $\Delta \theta (x)$ equals to $- \Delta k_x (x) / k_0$, where $\Delta k_x$ is the change in the projected wave vector of the ray in the lens plane and $k_0$ is the free space wave vector. The lens can focus rays emitted from a point source to another point, following the Gaussian lens formula $1 / u + 1 / v = \Delta \theta (x) / x = 1 / f$ ($u,\ v$: the distances from the object and the image to the lens in the $z$ direction; $f$: the focal length of the lens) \cite{jenkins1976fundamentals,born1999principles,hecht2017optics}. Therefore, we can obtain
\begin{equation}\label{eqn:1}
  \Delta k_x (x) = -k_0 x / f .
\end{equation}
Such a change in the wave vector can arise from a space-variant extra phase $\varphi(x)$ passing through the lens plane. The gradient of the phase change $\partial \varphi (x) / \partial x$ determines the wave vector change \cite{yu2011light,aieta2012aberration}, thus we have
\begin{equation}\label{eqn:2}
  \begin{gathered}
  \Delta k_x (x) = \partial \varphi (x) / \partial x = -k_0 x / f, \\
  \varphi (x) = - \frac{k_0 x^2}{2f} + \mathrm{Const}.
  \end{gathered}
\end{equation}
The constant can be neglected here. In other words, any optical element with a phase profile $\varphi (x) = - k_0 x^2/ 2f$ can function as a lens \cite{hecht2017optics,aieta2012aberration} because it can focus rays and image in the paraxial limit. A conventional lens introduces the phase profile by refraction through a bulk material. Alternatively, the required phase profile can be realized by specific space-variant resonances in a thin meta-lens.

\section{Reciprocal lenses: imaging elements based on ray shifting}
After reviewing the focusing and imaging mechanism of lenses, an interesting question arises: can electromagnetic waves be focused to produce images without bending rays? The answer is affirmative, as focusing and imaging can be realized by ray shifting. If each ray emitted by a point source can be
shifted in the $z$ direction by an identical distance $\tilde{f}$ after passing through some structure, an image of the source will be produced.

Equivalently, a ray with an incident angle $\theta$ can be shifted by a distance $\Delta x$ in the $x$ direction. In the paraxial limit, it will be $\Delta x (\theta) = \tilde{f} \cdot \theta$. Since we have $k_x \approx -k_0 \theta$, we can express $\Delta x$ with the in-plane wave vector $k_x$:
\begin{equation}\label{eqn:3}
    \Delta x (k_x) = - \tilde{f} k_x / k_0.
\end{equation}
The shifts of rays result from the extra phase gradient in the wave-vector space (momentum space) passing through the structure \cite{bliokh2013goos}, therefore we have
\begin{equation}\label{eqn:4}
  \begin{gathered}
    \frac{\partial\phi(k_x)}{\partial k_x} = - \Delta x (k_x) = \tilde{f} k_x / k_0, \\
    \phi(k_x) = \frac{\tilde{f} k_x^2}{2 k_0} + \mathrm{Const}.
  \end{gathered}
\end{equation}
We can conclude that any structure which provides an isotropic quadratic momentum-space phase modulation $\phi(\mathbf{k_\parallel}) \sim \left| \mathbf{k_\parallel} \right|^2$ can function as a ``lens'', which has a very different mechanism for focusing and imaging. Unlike lenses which are spatially variant, such a ``lens'' can be spatially homogeneous with no specific geometric center. As long as $u < \tilde{f} - l$ ($l$: the thickness of the ``lens''), the image of the source will appear on the opposite side of the reciprocal lens, and thus will be real. Moreover, the image of an object produced by this kind of ``lens'' will be upright as the directions of the rays remains unchanged.

We will refer to this class of ``lenses'' as ``reciprocal lenses'' for two reasons:

First, the distance between the source and the image is a constant, leading to a different lens formula,
\begin{equation}\label{eqn:5}
    u + v = \tilde{f}.
\end{equation}
The form of this lens formula (Eq. \ref{eqn:5}) is the reciprocal of the well-known Gaussian lens formula $1 / u + 1 / v = 1 / f$. The phase distribution function (Eq. \ref{eqn:4}) is also the reciprocal of that of lenses (Eq. \ref{eqn:2}). Note that $\tilde{f}$ in Eq. \ref{eqn:5} plays a similar role to the focal length $f$ if we compare the formula with the Gaussian one. Thus, we can define $\tilde{f}$ as the focal length of a reciprocal lens.

Second and most importantly, reciprocal lenses are based on ray shifting effect as we derived, which is the reciprocal counterpart effect of ray bending. Rather than changing $\theta(x)$ -- the propagation directions of rays at different positions, reciprocal lenses modulate $x(\theta)$ -- the positions of rays with different incident angles. The roles of $x$ and $\theta$ (or $k_x$) are exchanged.

\begin{figure*}[htpb]
\includegraphics[scale=0.9]{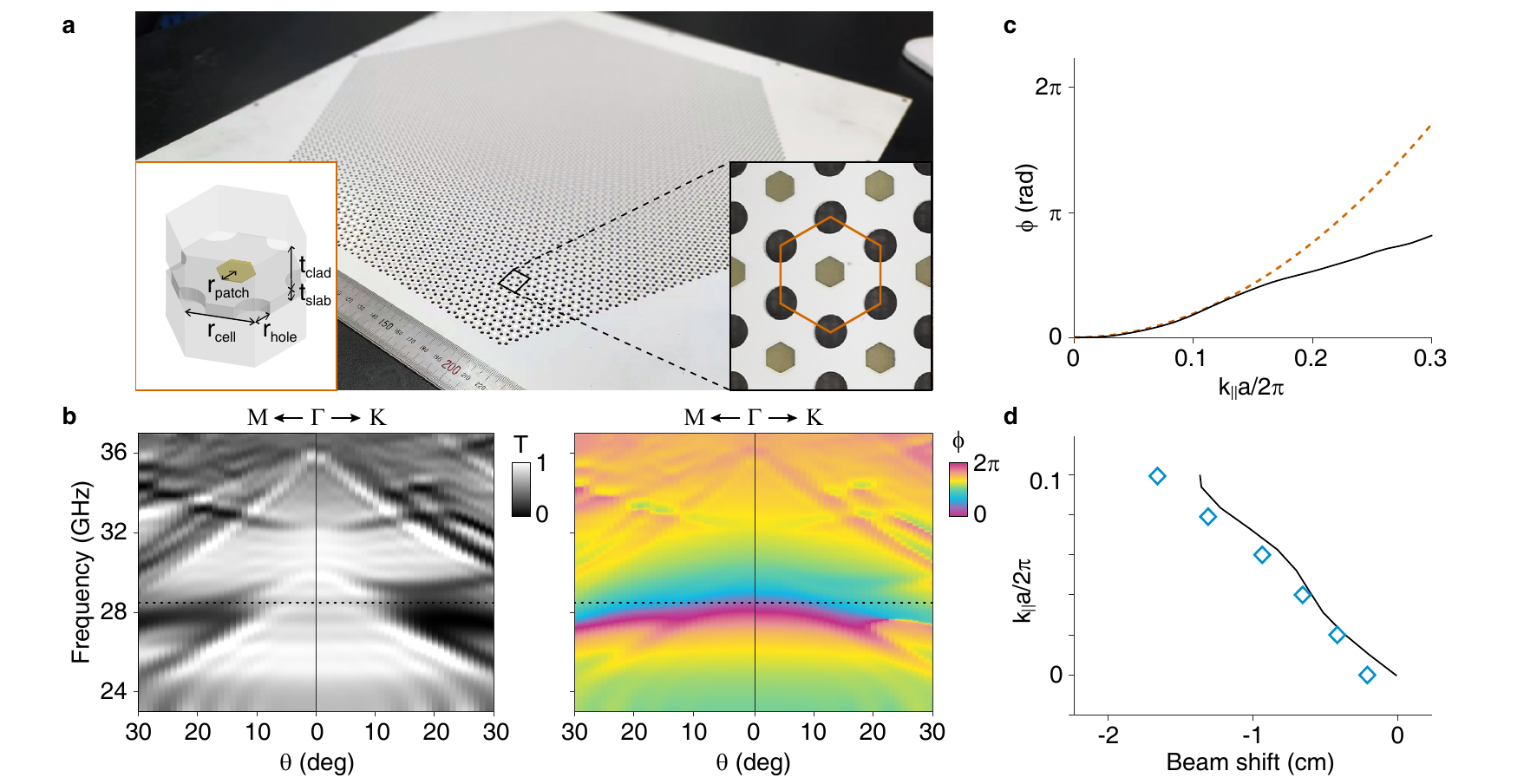}
\centering
\caption{Characteristics of the designed ultra-thin reciprocal lens. (a) A photo of the core layer of the designed reciprocal lens comprising a honeycomb array of circular holes etched on a printed circuit board (PCB), whose permittivity is 6.15 and loss tangent is 0.002 (Arlon TC600). A hexagonal metallic patch is printed at the center of each honeycomb unit cell. The structured PCB acts as a photonic crystal slab and is clad by two background layers (Arlon DiClad 880, permittivity: 2.2; loss tangent: 0.0009). Left inset: the schematics of the unit cell; right inset: a zoomed-in photo of the structure. (b) The angle-resolved transmittance and transmissive phase change spectra of the structure (with circularly co-polarized input and output). (c) The phase change curve along the $\Gamma$-K direction at the working frequency 28.5 GHz. In a specific wave vector (angle) range, the phase induced by the reciprocal lens (the solid curve) follows the quadratic rule we desired, of which the fitted quadratic dependence is plotted as the dashed curve. (d) The beam shifts for different incident angles corresponding to the momentum-space phase gradient. The shifts of the beam centroids for different incident angles (blue diamonds) follow the displacements calculated from the momentum-space phase gradients (black curve). It can be seen that the beam shift has the expected linear dependence on the in-plane wave vector.}
\label{fig:2}
\end{figure*}

Conceptually, slabs supporting negative refraction can be viewed as examples of reciprocal lenses \cite{kosaka1998superprism,pendry2000negative,smith2004metamaterials,fang2005sub,eleftheriades2005negative,engheta2006metamaterials,liu2007far,yao2008optical,valentine2008three,wong2017optical}, as illustrated in Fig. \ref{fig:1}. An arbitrary ray will be bent twice on passing through the slab, effectively resulting in a shift which corresponds to the $z$-direction propagation phase gained by the ray. The focal length $\tilde{f}$ of such a slab is $l\left( 1 - n_0/n_1\right)$ ,where $l$ is the thickness of the slab, $n_0$ is the background refractive
index, and $n_1$ is the refractive index of the slab. When a reciprocal lens based on negative refraction is combined with the evanescent-wave-amplification effect, sub-wavelength imaging can be realized in the near field \cite{pendry2000negative,fang2005sub,liu2007far}. However, there are difficulties in realizing a bulky slab with negative refraction \cite{yao2008optical,valentine2008three}. More importantly, realizing reciprocal lenses based on negative refraction still involves ray bending at the interfaces, and the ray shifting relies on the propagation phase given by the bulk dispersion of a thick slab.

\section{Realization of an ultra-thin reciprocal lens}

In this work, we eliminate the bulky structure and to realize a reciprocal lens based on ray shifting effect only. Instead of leveraging the propagation phase, we apply guided resonances supported by a two-dimensional (2D) photonic crystal (PhC) slab \cite{yablonovitch1987inhibited,john1987strong,johnson1999guided,fan2002analysis,guo2020squeeze,guo2021structured,long2022polarization} as a non-local meta-surface to induce the momentum-space phase modulation, which allows the resulting reciprocal lens to be ultra-thin. When the incident and outgoing polarization state of the plane waves passing through the PhC slab are the same (co-polarization condition), it can be shown that a specific photonic band of guided resonances could give approximately a $\left|\mathbf{k_\parallel}\right|^2$-dependent resonant phase modulation to the transmitted waves with $\mathbf{k_\parallel} \approx \mathbf{0}$ \cite{guo2020squeeze,long2022polarization}. The derivations and discussions of the transmissive phase induced by a 2D PhC are detailed in Appendix A. If the phase modulation meets the requirements that it is positively quadratic and sufficiently isotropic, the PhC slab will behave like a reciprocal lens.

We designed and fabricated a 2D-PhC-slab reciprocal lens operating in the microwave range using printed circuit boards (PCBs). As shown in Fig. \ref{fig:2}(A), a one-millimeter-thick PCB (permittivity: 6.15; loss tangent: 0.002) structured with circular holes and printed with hexagonal metallic patches acts as the core 2D-PhC-slab layer. The holes are arranged periodically in a honeycomb array, and the distance between two neighboring holes $r_\mathrm{cell}$ is 3.648 mm. In each unit cell, there is a patch in the middle of the six holes. Such a lattice is $C_{6v}$-symmetric. The structured core layer supports both transverse-electric-like (TE-like) and transverse-magnetic-like (TM-like) guided resonances as a result of the Bragg scattering of the periodic array, which is necessary in our design for momentum-space phase modulation. The core layer is clad by two background layers (thickness: 4.725 mm; permittivity: 2.2; loss tangent: 0.0009), taking the possibility of multi-layer stacking into consideration.

In our experiments, the input and output waves are chosen to be both left-handed (or right-handed) circularly polarized. Combining the circular symmetry of the polarization states and the $C_{6v}$ symmetry of the structure, the system is $C_6$-symmetric, which guarantees good isotropic phase modulation provided by the reciprocal lens in the momentum space. It can be seen from the measured angle-resolved phase spectra [right panel of Fig. \ref{fig:2}(b)] that the momentum-space phase modulation of the reciprocal lens is nearly the same for the two different high-symmetric directions $\Gamma$-K and $\Gamma$-M near $\mathbf{k_\parallel} = \mathbf{0}$. Note that the momentum-space-variant transmittance at the working frequency also affects the quality of focusing and imaging. In order for the transmittance to approach unity and to have a large enough phase span, we tune the radii of the holes ($r_\mathrm{hole}$: 1.113 mm) and the patches ($r_\mathrm{patch}$: 1.094 mm) respectively. Consequently, one pair of TE-like degenerate resonances overlap with another pair of TM-like resonances in the spectrum, leading to a high co-polarized transmittance \cite{sm,chen2018huygens,liu2021ways,eleftheriades2022prospects}. Detailed discussions about enhancing the transmittance by overlapping degenerate resonances can be found in Appendix B \& C. As shown in the left panel of Fig. \ref{fig:2}(b), the measured transmittance is high near $\mathbf{k_\parallel} = \mathbf{0}$ over a large spectral range.

We pick the working frequency to be $28.5$ GHz [marked by a dashed line in Fig. \ref{fig:2}(b)], where the momentum-space phase distribution best fits the positive quadratic dependence and gives the greatest focal length. The $\mathbf{k_\parallel}$ phase modulation function along the $\Gamma$-K direction at the working frequency is measured and plotted in Fig. \ref{fig:2}(c). As mentioned, the phase induced by the structure follows the quadratic rule over a specific range of $\mathbf{k_\parallel}$. We find by curve fitting that the focal length $\tilde{f}$ at $28.5$ GHz is about $7$ cm, which is nearly 7 times the wavelength. In addition, we measured the shift of a Gaussian beam (waist radius $\sim4.57$ cm) with different central wave vectors in the sample plane, which is plotted in Fig. \ref{fig:2}(d). The incident in-plane wave vectors are chosen to be in the $\Gamma$-K direction. The beam shifts confirm that the momentum-space phase modulations induced by the designed structure can shift rays in the expected way as illustrated in Fig. \ref{fig:1} and described by Eq. \ref{eqn:4}, i.e., it can work as a reciprocal lens.

\begin{figure}[htpb]
\includegraphics[scale=0.9]{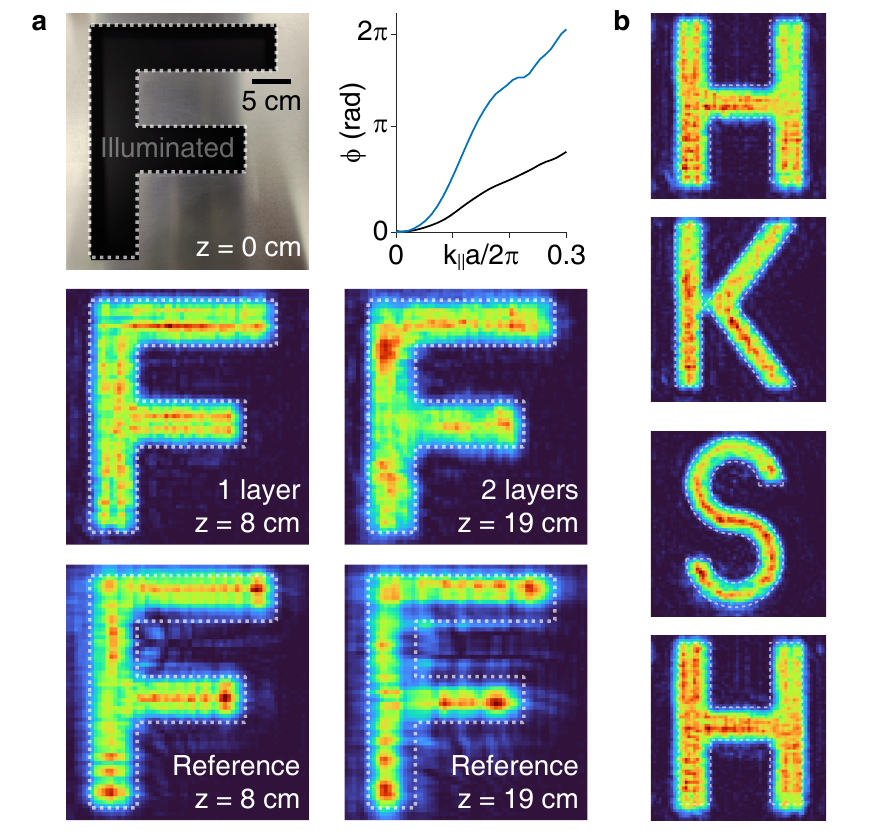}
\centering
\caption{Imaging by reciprocal lens. (a) The ``$z = 0$ cm'' photo shows the object, which is a metallic plate with an ``F''-shaped slot cut in it. A horn antenna is used to illuminate the slot. With one layer of our designed reciprocal lens applied at $z = 3$ cm, an upright real image of the ``F''-slot can be seen clearly at z = 8 cm. In comparison, only a diffracted pattern can be seen without the reciprocal lens. Further, two layers of the designed reciprocal lens are cascaded to increase the focal length. The focal length can be doubled, as shown in the second plot of the right panel. (b) Several different shapes of objects are also used to demonstrate the imaging effect.}
\label{fig:3}
\end{figure}

As shown in Fig. \ref{fig:3}, we demonstrated the imaging effect of the designed ultra-thin reciprocal lens by measuring the transmitted field pattern directly. We use a metallic plate with an ``F''-shaped slot cut in it as the imaging object, which is shown in the left panel of Fig. \ref{fig:3}(a). The object is set at a position marked as $z = 0$, and the slot is illuminated uniformly by a horn antenna. First, one layer of the designed reciprocal lens is placed at $z = 3$ cm. Consequently, we observe a clear upright real image of the object ``F''-slot at $z = 8$ cm with sharp boundary profiles. The image is the same size as the object. For comparison, a diffracted pattern of the object is seen if the reciprocal lens is removed.

In addition, reciprocal lenses can be stacked to increase the focal length. By stacking multiple identical reciprocal lenses, the momentum-space phase modulation may be enlarged additively, giving a focal length several times the original one. To demonstrate this fact, we applied two layers of the designed reciprocal lens placed at $z = 3$ cm, stacked immediately on top of each other. We note that the thickness of the cladding layers has been chosen to reduce the interaction between the photonic crystal slabs in the stacked reciprocal lens. If we simply stack the photonic crystal slabs on top of each other without the spacers, the interaction between the resonances of adjacent photonic crystal slabs may affect the total transmittance and transmissive phase so that the focal length will not be additive and may degrade the performance of the stacked reciprocal lens. From the measured momentum-space phase distribution [the first plot in the right panel of Fig. \ref{fig:3}(a)], we find that the focal length is nearly doubled to about 19 cm. We confirmed the increased focal length by measuring the field profile as well. The image produced at $z = 19$ cm is of good quality, while the field without the stacked reciprocal lens at $z = 19$ cm is diffracted more strongly. We also used several different shapes of objects, such as ``H'', ``K'' and ``S'', to further test the imaging effect of the one-layer reciprocal lens. All the shapes are well imaged, as shown in Fig. \ref{fig:3}(b). The acute angles of ``K'' and curved edges of ``S'' are well reproduced in the images, indicating that there is no constraint on the shape of objects. The two images of the ``H''-shaped slot which are obtained from separate imaging tests appear to be almost identical, proving the imaging repeatability. The detailed experiment setups can be found in Appendix D.

\section{Conclusions}
In short, we proposed and demonstrated the concept of the PhC-slab-based ``reciprocal lens'', imaging components which produce images without bending rays. Distinctly different from conventional lenses and meta-lenses, the images produced by our PhC-slab-based reciprocal lens are upright but real, and the object-image distance is a constant. Our reciprocal lens is composed of an ultra-thin periodic structure, and has no geometric center. As such, it is not only simple to fabricate but also easy to integrate, and it can image multiple objects simultaneously due to non-locality. PhC-slab-based reciprocal lenses complement the family of optical elements and may have potential applications in various photonic devices. They also provide a new ray-shifting paradigm for imaging and wavefront-reshaping.

\section*{Acknowledgments}
\textbf{Funding:} This work is supported by RGC through grant AoE/P-502/20, the Croucher Foundation (CAS20SC01), and also by the China National Key Basic Research Program 2018YFA0306201. L. S. is further supported by Science and Technology Commission of Shanghai Municipality (20501110500 and 21DZ1101500).

W. L., J. C., and T. L. contributed equally to this work.

%

\end{document}